\documentclass[aps,prd,secnumarabic,nobibnotes,twocolumn,superscriptaddress]{revtex4-1}
\usepackage{amsfonts}
\usepackage{mathrsfs}
\usepackage{amsmath}
\usepackage{color}
\usepackage{natbib}
\usepackage{graphicx}
\usepackage{bm}
\usepackage{amssymb}
\usepackage{xspace}
\usepackage{epstopdf}
\usepackage{dcolumn}
\usepackage{multirow}
\usepackage[colorlinks=true, letterpaper=true, pdfstartview=FitV, linkcolor=blue, citecolor=blue, urlcolor=blue]{hyperref}
\usepackage{wrapfig}

\makeatletter

\newcommand{\Rmnum}[1]{\expandafter\@slowromancap\romannumeral #1@}
\makeatother

\begin{document}
\title{Topological Phase with Critical-Type Nodal Line State in Intermetallic CaPd}

\author{Guodong Liu}
\address{School of Materials Science and Engineering, Hebei University of Technology, Tianjin 300130, China}

\author{Lei Jin}
\address{School of Materials Science and Engineering, Hebei University of Technology, Tianjin 300130, China}

\author{Xuefang Dai}
\address{School of Materials Science and Engineering, Hebei University of Technology, Tianjin 300130, China}

\author{Guifeng Chen}
\address{School of Materials Science and Engineering, Hebei University of Technology, Tianjin 300130, China}

\author{Xiaoming Zhang}
\email{zhangxiaoming87@hebut.edu.cn}
\address{School of Materials Science and Engineering, Hebei University of Technology, Tianjin 300130, China}

\begin{abstract}
In recent years, realizing new topological phase of matter has been a hot topic in the fields of physics and materials science. Topological semimetals and metals can conventionally be classified into two types: type-I and type-II according to the tilting degree of the fermion cone. Here, it is the first time to report a new topological metal phase with the critical-type nodal line between type-I and type-II nodal line. The critical-type nodal line shows a unique nontrivial band crossing which is composed of a flat band and a dispersive band and leads to a new fermionic state. We propose intermetallic CaPd can be an existing topological metal for the new fermionic state, characterized with critical-type nodal line in the bulk and drumhead band structure on the surface. Our work not only promotes the concept of critical-type nodal line, but also provides the material realization to study its exotic properties in future experiments.
\end{abstract}
\maketitle

\section{Introduction}
Since the theoretical prediction and experimental confirmation of topological insulators~\cite{1,2,3,4,5}, increasing interests in topological materials have been motivated not only by the developing requirements of fundamental physics but also by their potential technological applications. Among various topological phases, topological semimetals/metals (TMs) have become the main research focus in recent years~\cite{6,7,8,9,10,11,12,13}. TMs are characterized by symmetry-protected band crossings in the low-energy band structures. According to the dimensionality of band crossings, TMs are usually classified into three types: (1) topological nodal point semimetals/metals (TNPMs) such as Weyl and Dirac semimetals/metals~\cite{8,9,10,11,12,13,14,15,16,17,18}; (2) topological nodal line semimetals/metals (TNLMs)~\cite{19,20,21,22,23,24,25,26,27,28,29,30,31,add4,add5}; and (3) topological nodal surface semimetals/metals (TNSMs)~\cite{32,33,34,add3}. Their band crossings form zero-dimensional nodal point, one-dimensional nodal line, and two-dimensional nodal surface in TNPMs, TNLMs and TNSMs, respectively. Among these TMs, TNLMs have received increasing interests currently. They are firstly predicted in several families of materials in theory~\cite{19,20,21,22,23,28,29,30,31,32,33}, then increasing experimental confirmation has been achieved by angle-resolved photoemission spectroscopy (ARPES)~\cite{35,36,37,38}. Recently, novel transport, magnetic and optical properties are also observed in TNLMs~\cite{39,40,41,42}.

	Depending on the slope of band dispersion in the momentum-energy space, TMs are previously termed as two types, namely type-I and type-II TMs. In type-I TMs, the bands manifest conventional conical dispersions [see Fig. 1(a)], with electron and hole regions well separated by the energy. In type-II TMs, the spectrum is completely tipped-over [see Fig. 1(c)], which can give rise to the coexistence of electron and hole states at a given energy~\cite{43,44,45,46}. The physical properties of type-I and type-II TMs are drastically different~\cite{47,48,49,50,51}. Quite recently, the theory of the critical state between type-I and type-II TMs has also been established in Dirac/Weyl semimetals~\cite{52,53,54}. As shown in Fig. 1(b), the band structure of the critical-type TMs is characterized by the band crossing between a horizonal band and a dispersive band. Most importantly, the critical-type Dirac/Weyl fermions are predicted to show unique physical properties, such as Dirac/Weyl line Fermi surface, critical chiral anomaly and unexpected topological transition under magnetic field~\cite{52,53,54}. For critical-type TMs, current study mostly stays at the theoretical model/concept stage, and quite rare candidate materials have been identified (except for the critical-type Dirac nodal point proposed in Zn$_2$In$_2$S$_5$ compound~\cite{54}). Especially, up to date, the nodal line version of critical-type band crossing has not been proposed, and it is urgent to identify an existing material that hosts critical-type nodal line state.

\begin{figure}
\includegraphics[width=8.8cm]{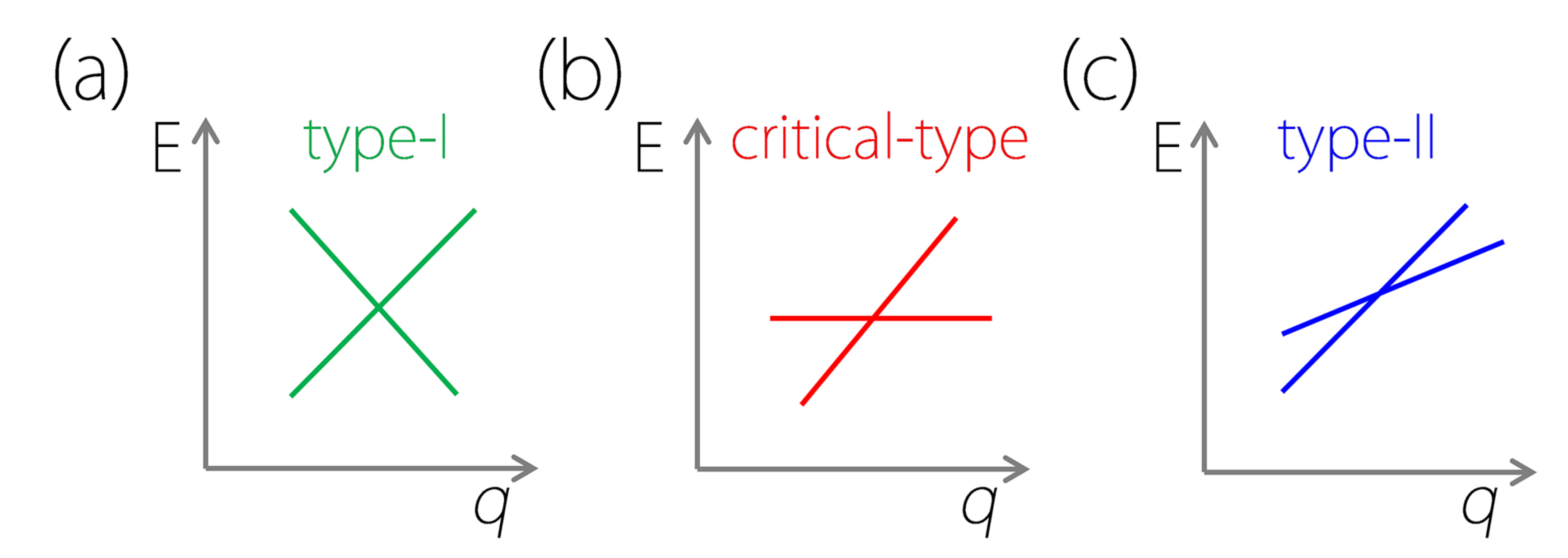}
\caption{Schematic figure of (a) type-I, (b) critical-type and (c) type-II band dispersions in the momentum-energy space.
\label{fig1}}
\end{figure}

In this work, we will fill in this research gap by identifying the critical-type nodal line state in intermetallic material CaPd. Based on first-principles calculations, we find that there exists a closed nodal line centering X point in the k$_{x/y/z}$=$\pi$ plane in CaPd. Interestingly, the nodal line is formed by the crossing of a nearly flat band and a dispersive band. The unique slope of band crossing identified here is drastically different from conventional type-I and type-II nodal lines in previous examples and is considered as a critical-type nodal line state which is possible to produce new fermions. As a clear signature of nodal line materials, the drumhead surface state is observed in CaPd. We also find that, the formation of the critical-type nodal line is extremely material-specific and the candidate material is quite in scarcity. Our work not only proposes the concept of critical-type nodal line state in theory, but also provides an existing material to study its novel properties in future.

\section{Methods}

The first-principles calculations were performed in the framework of density functional theory (DFT) by using the Vienna ab-initio simulation package (VASP)~\cite{55,56}. For ionic potentials, we used the generalized gradient approximation (GGA) of Perdew-Burke-Ernzerhof (PBE) method~\cite{57}. The cutoff energy was adopted as 500 eV, and the Brillouin zone (BZ) was sampled with $\Gamma$-centered kmesh of 15$\times$15$\times$15 for both structural optimization and self-consistent calculations. The energy convergence criteria was chosen as 10$ ^{-7}$ eV. The surface states were studied by using software WannierTools~\cite{58} based on the method of maximally localized Wannier functions~\cite{59,60}.

\section{Results and discussions}

\begin{figure}
\includegraphics[width=8.8cm]{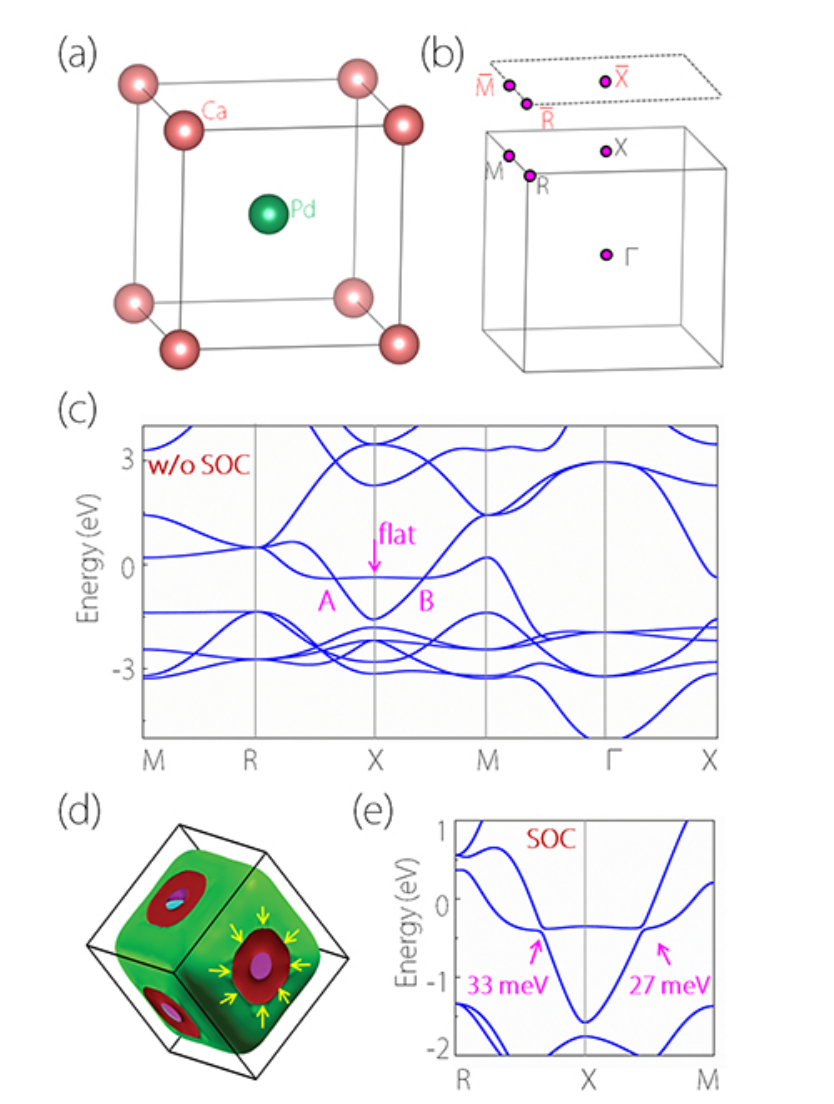}
\caption{(a) Crystal structure of intermetallic CaPd. (b) The bulk and surface Brillouin zone with high-symmetry points labelled. (c) Electronic band structure of CaPd along high-symmetry lines without considering SOC. (d) Isoenergy surface of CaPd at E= E$_F$ - 0.38 eV. The crossing boundary between the two hole pockets is pointed by the yellow arrows. (e) Electronic band structure of CaPd along R-X and X-M paths with considering SOC. The sizes of SOC-induced gaps are indicated in the figure.
\label{fig2}}
\end{figure}

	Intermetallic CaPd is an existing material, firstly prepared by Mendelsohn \emph{et al.} in 1973 ~\cite{add2}. The preparation of CaPd follows two processes: (1) pressing a 5:3 mixture of pure calcium and palladium into a pellet; (2) heating the pellet at $\sim$900$^{\circ}$C for 12 h in the argon atmosphere. Later on, Iandelli \emph{et al.}~\cite{61} and Hawanga \emph{et al.}~\cite{62} have also synthesized high-quality crystal samples of CaPd by using similar method. The X-ray results indicate that CaPd has a cubic structure with the CsCl-ordered crystal type. As shown in Fig. 2(a), in the unit cell, Ca and Pd atoms occupy \emph{1a} (0, 0, 0) and \emph{1b} (0.5, 0.5, 0.5) Wyckoff sites, respectively. The optimized lattice constant yields to be 3.534 \AA, in good accordance with the experimental one (3.518 \AA) ~\cite{61,62}.

The electronic band structure of intermetallic CaPd without spin-orbit coupling (w/o SOC) is shown in Fig. 2(c). It exhibits a metallic band structure, and near the Fermi energy the bands form two crossing points: crossing-A on the X-R path and crossing-B on the X-M path. Quite interestingly, the two crossing points are composed of a nearly flat band and a dispersive band, making both crossing-A and crossing-B are critical-type nodal points.

	Noticing the CaPd system reserves both time reversal (\emph{T}) and spatial inversion (\emph{P}) symmetries, crossing-A and crossing-B cannot be isolate nodal points but belong to nodal line or other nodal structures, because under the protection of \emph{P} and \emph{T} symmetries the spinless Hamiltonian is always real valued~\cite{19}. To verify this, we first calculate the isoenergy surfaces by setting the energy level near the band crossing points (-0.38 eV). As shown in Fig. 2(d), we find the isoenergy surfaces are formed by two hole pockets which cross with each other. The crossing boundary between the hole pockets clearly manifests a closed loop in the k$_{x/y/z}$=$\pi$ plane, as shown by the arrows in Fig. 2(d). This indicates crossing-A and crossing-B most likely belong to a nodal line in the k$_{x/y/z}$=$\pi$  plane. So we make a careful investigation to band structure in the k$_{z}$=$\pi$ plane. As shown in Fig. 3(a), crossing-A and crossing-B indeed reside on a nodal line, which centers X point in the k$_{z}$=$\pi$ plane. Because of the cubic symmetries, there also possess nodal line in the equivalent k$_{x}$=$\pi$ and k$_{y}$=$\pi$ plane. In fact, the nodal line is constrained to lie in the k$_{x/y/z}$=$\pi$ plane by the mirrior symmetry \emph{M$_{x,y,z}$}. The nodal line locates in the mirror-invariant plane, and it also enjoys the protection of \emph{M$_{x,y,z}$}, as the two crossing bands possess opposite \emph{M$_{x,y,z}$}-eigenvalues.

\begin{figure}
\includegraphics[width=8.8cm]{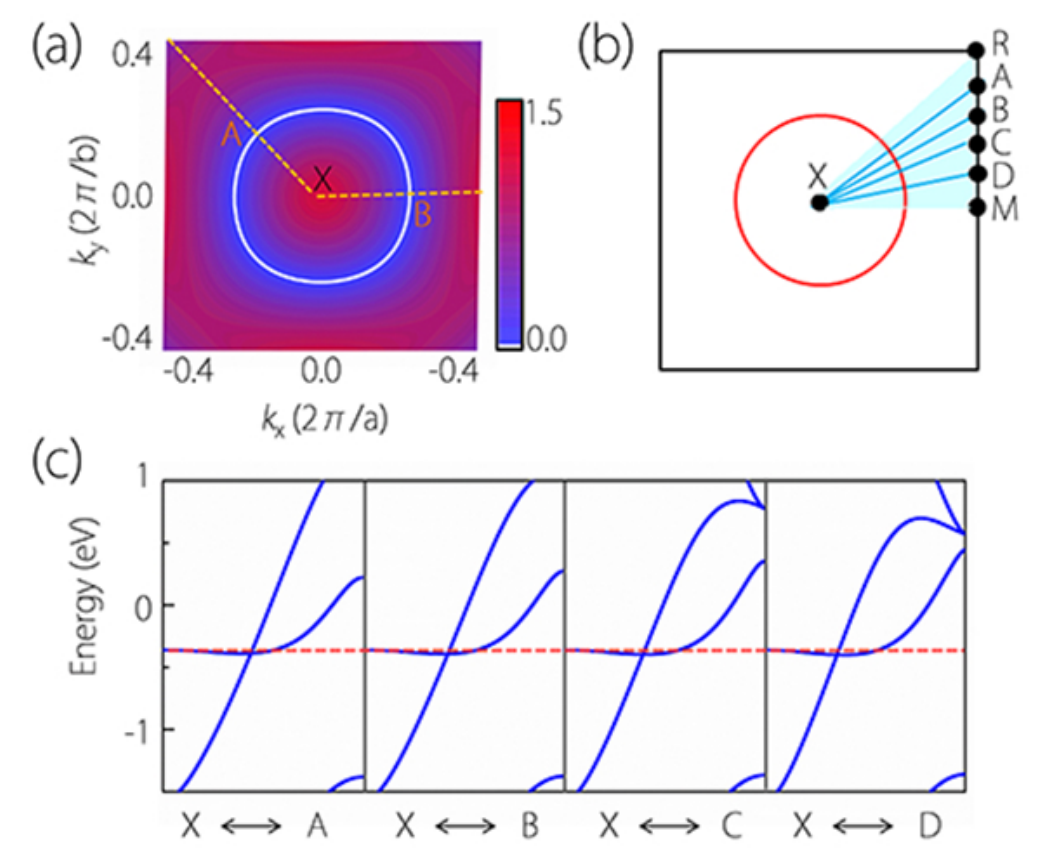}
\caption{(a) The shape of nodal line in the k$_{z}$=$\pi$ plane. The positions of crossing-A and crossing-B are indicated in the figure.  (b) The selected \emph{k} paths (X-A, X-B, X-C and X-D) through the nodal line. The points A, B, C, and D are equally spaced between M and R. (c) Electronic band structures of CaPd along X-A, X-B, X-C and X-D paths. In (c), the horizontal dashed line serves as a guide of eye.
\label{fig3}}
\end{figure}

	When SOC is absent, the nodal line in CaPd is stabilized by the coexistence of \emph{P} and \emph{T} symmetries in the system. Similar to most nodal line materials, SOC will gap the nodal line in CaPd. As shown in Fig. 2(e), our results indicate that in CaPd the SOC-induced gaps at crossing-A and crossing-B are $\sim$30 meV, which are comparable or smaller than those in typical nodal line materials, such as ZrSiS ($>$20 meV)~\cite{36}, TiB$_2$ ($>$25 meV)~\cite{28}, Mg$_3$Bi$_2$ ($>$36 meV)~\cite{30}, Cu$_3$NPd ($>$ 60 meV)~\cite{22,23}, CaAgBi ($>$80 meV)~\cite{63}, and so on. Considering the nodal line materials with similar size of SOC-induced gaps including ZrSiS~\cite{36,37,39}, TiB$_2$~\cite{64}, and Mg$_3$Bi$_2$~\cite{65} have been experimentally verified, the nodal line signature in CaPd is also promising to be detected in experiments.

As shown in Fig. 2(c), on the nodal line crossing-A and crossing-B manifest critical-type band crossing. Here we make further investigation to the band crossing on other parts of the nodal line. Noticing both fourfold rotation \emph{C$_4$} symmetry and mirror \emph{M$_z$} symmetry preserve in the k$_{z}$=$\pi$ plane, 1/8 part of the first BZ can reflect the nature of the whole. Besides the X-M and X-R paths, we have selected four paths in the region, namely X-A, X-B, X-C, and X-D [see Fig. 3(b)]. The band structures on the four paths are shown in Fig. 3(c). We find that all the paths possess nearly critical-type band crossings, indicating the truth of critical-type signature on the whole nodal line.

It is worth noticing that, the concept of critical-type nodal line state has never been proposed before. Considering that critical-type Dirac/Weyl points show different topological properties (such as critical chiral anomaly and interesting topological transition under magnetic field) with traditional type-I and type-II nodal points, critical-type nodal lines are also expected to possess these unique characteristics, because each point on the nodal line can be viewed as a critical-type nodal point in the transverse dimensions perpendicular to the line. Moreover, critical-type nodal line should also possess distinct properties with critical-type nodal points because of their diverse dimensionality of band crossing. Such distinct properties are quite promising to be identified from transport experiments. One observation is the phase shift of the Shubnikov-de Haas oscillation. For example, previous transport studies on nodal line material ZrSiS found that the phase shift in the higher frequency component can diverse from 0 to $\pm$1/8 under different conditions~\cite{add6,add7}, being different from the fixed phase shift in a given Dirac/Weyl material. Especially, based on the general rule by Li \emph{et al.}~\cite{add1} established quite recently, the phase shifts of quantum oscillations in nodal line materials can show rich results depending on
the magnetic field directions, carrier types, and configurations of the Fermi surface, quite different from those from normal electrons and Dirac/Weyl fermions. For the proposed critical-type nodal line, it is promising to observe both features of critical-type nodal point (such as the critical chiral anomaly) and of nodal lines (such as rich phase shifts of quantum oscillations from their nontrivial Berry phases).

\begin{figure}
\includegraphics[width=8.8cm]{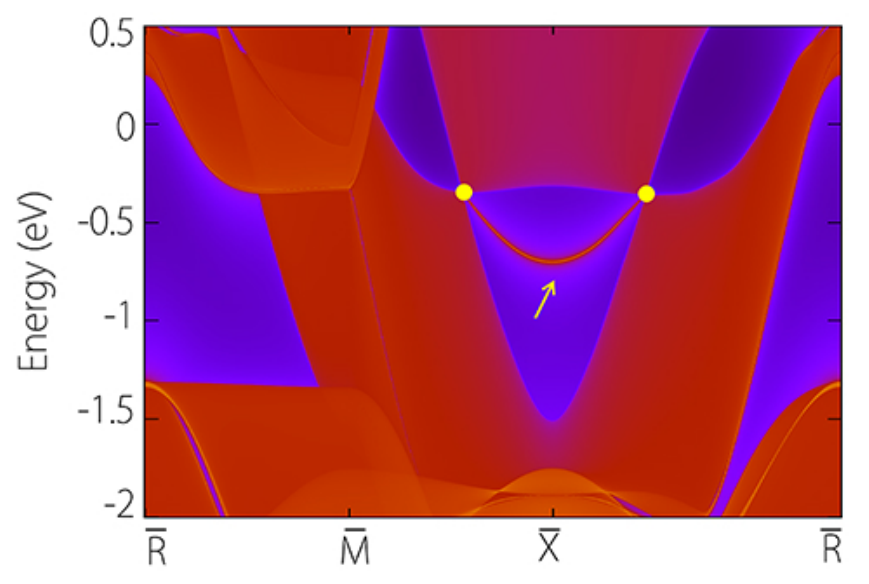}
\caption{(a) Projected spectrum for the Ca-terminated (001) surface of intermetallic CaPd. The yellow dots mark the position of projected bulk band-crossing points, and the yellow arrow indicates the drumhead surface states.
\label{fig4}}
\end{figure}

	One of the most representative signature of a nodal line state is the existence of drumhead surface states~\cite{19}. For intermetallic CaPd, we show the  Ca-terminated (001) surface spectrum in Fig. 4. Like type-I and type-II nodal lines, we find critical-type nodal line also manifests drumhead surface state emanated from the bulk nodal points, as pointed by the yellow arrow in Fig. 4. Because of the absence of particle-hole symmetry, the surface band is not perfectly flat but possesses an energy dispersion of about 300 meV. Similar dispersive surface states are also observed in other nodal line materials~\cite{19,20,21,22}. The existence of drumhead surface states in CaPd can be readily detected by surface sensitive probes such as ARPES and scanning tunneling microscopy.

\begin{figure}
\includegraphics[width=8.8cm]{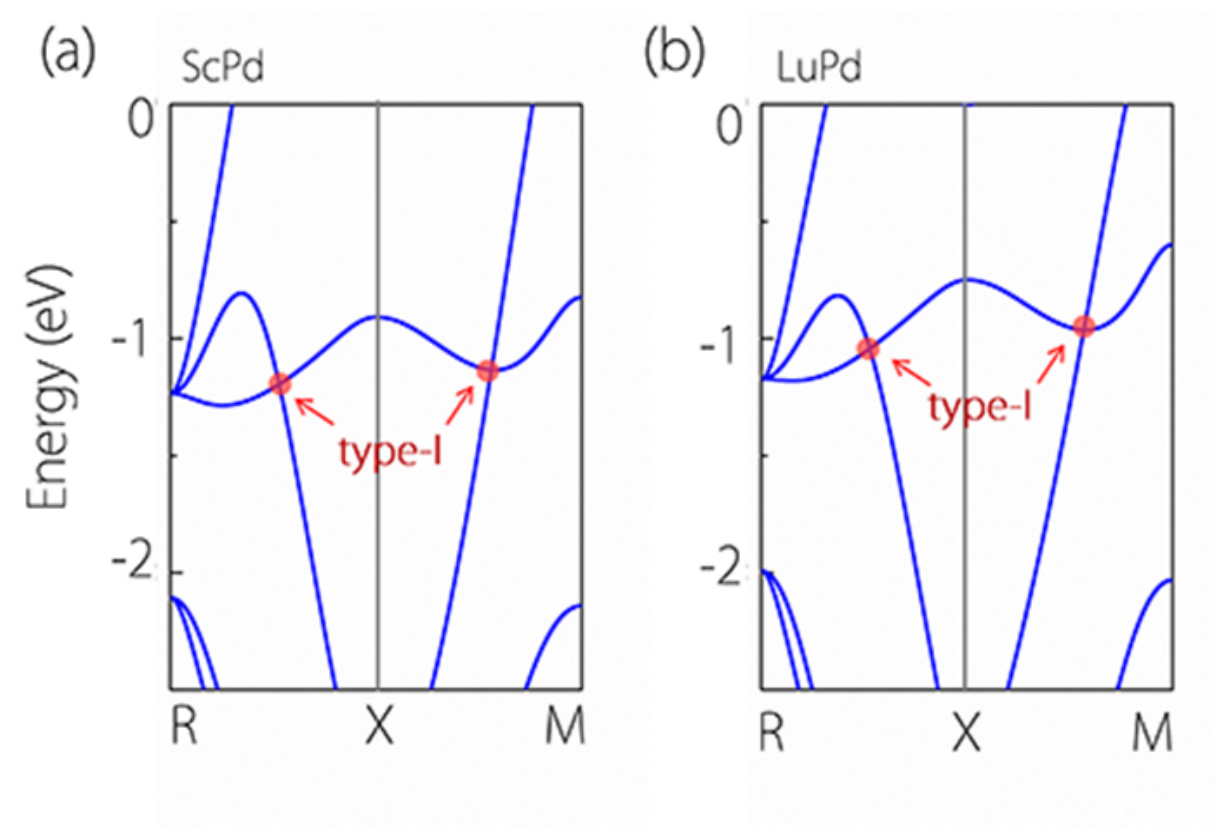}
\caption{Electronic band structures along the R-X and X-M paths for (a) ScPd and (b) LuPd. The crossing points in (a) and (b) show type-I band dispersion.
\label{fig5}}
\end{figure}

	Finally, to further understand the nature of the nodal line state in CaPd, we construct a low-energy effective Hamiltonian at X point. The symmetry analysis indicates the two bands belong to irreducible representations \emph{A$_{1g}$} and \emph{A$_{2u}$} of the\emph{ D$_{4h}$} symmetry, respectively. Using them as basis states, the $2\times 2$ effective Hamiltonian up to the quadratic order can be described as :

\begin{equation}\label{FNRm}
\mathcal{H}=\left[
              \begin{array}{cc}
                h_{11} & h_{12} \\
                h_{21} & h_{22} \\
              \end{array}
            \right],
\end{equation}
with,
\begin{equation}
h_{ii}=A_{i}(k_{x}^{2}+k_{y}^{2})+B_{i}k_{z}^{2}+M_{i}
\end{equation}
\begin{equation}
h_{12}=-h_{21}=iCk_{z}
\end{equation}

From the Hamiltonian, we find the non-diagonal terms will vanish when \emph{k$_z$}=0 (notice, here the wave vector \emph{k} is expanded from X point). This indicates that the band-crossing in the k$_{x/y/z}$=$\pi$ plane will produce a nodal line, which is consistent with the results from DFT calculations. In the Hamiltonian, \emph{A$_i$}, \emph{B$_i$}, \emph{M$_i$} (\emph{i}=1,2) and \emph{C} are material-specific coefficients, which determine the details of band structure near X point. In CaPd, these model parameters can be obtained by fitting the DFT band structure. The fitting results yield to be $A_1 = 0.09$ eV$\cdot${\rm \AA}$^2$, $A_2 = 7.25$ eV$\cdot${\rm \AA}$^2$, $B_1 = 17.22$ eV$\cdot${\rm \AA}$^2$, $B_2 = -22.31$ eV$\cdot${\rm \AA}$^2$, $C = 15.17$ eV$\cdot${\rm \AA}$^2$, $M_1 = -0.28$ eV, and $M_2 = -1.46$ eV. The small value of $A_1$ produces the nearly non-dispersive valence band around X point, which could be advantageous for achieving the critical nodal line. It should be mentioned that an extremely small $A_1$ and specific energy variation may lead to possible quantum phase transition from  paramagnetic to ferromagnetic induced by the electron-electron interaction, which has been well described in the critical point between type-I and type-II TNLMs by He\emph{ et al.}~\cite{add8} quite recently. However, we note that for CaPd, the critical nodal line is stable, as there is no appearance of ferromagnetic order in  our DFT calculations. From the Hamiltonian, we can also find that, beside the protection of specific symmetries, the formation of a critical-type nodal line is extremely material-specific. To show this point, we choose other two intermetallic compounds namely ScPd and LuPd as examples. Both ScPd~\cite{66} and LuPd~\cite{67} are existing materials with the same CsCl-type structure. The electronic band structures of ScPd and LuPd are shown in Fig. 5(a) and (b). We find that there also exist band crossing points on the R-X and X-M paths in ScPd and LuPd. Under the same protection mechanism with CaPd, the crossing points in ScPd and LuPd also belong to nodal lines centering X point in the k$_{x/y/z}$=$\pi$ plane. However, the nodal lines in ScPd and LuPd possess conventional type-I band dispersion, drastically different from the critical-type nodal line state in CaPd. Such high material-dependence indicates the number of critical-type nodal line semimetal is quite limited in real materials.

Before closing, we want to point out that, in the ground state the critical-type nodal line in CaPd does not close to the Fermi level (at -0.38 eV), which will greatly hamper its experimental identification (such as from transport experiments). Fortunately, we find the position of nodal line in CaPd can be tuned near the Fermi level by strain and proper elemental doping, with the critical band-crossing signature nearly unaffected (see APPENDIX A). Therefore, the proposed CaPd-based materials can serve as excellent platform to study the novel properties of the newly proposed critical-type nodal line state.

\section{Summary}

In summary, we have proposed a new topological phase with nearly critical-type nodal line state in intermetallic CaPd. The critical-type nodal line is the transition state from conventional type-I to type-II nodal lines, and is formed by the crossing of a flat band and a dispersive band in the momentum-energy space. In intermetallic CaPd, the critical-type nodal line resides on the k$_{x/y/z}$=$\pi$ plane, protected by both mirror symmetry and the coexistence of \emph{P} and \emph{T} symmetries. Clear drumhead surface state originating from critical-type nodal line is observed in CaPd, which is quite promising to be detected in experiments. We find that the existence of critical-type nodal line is quite material-specific, so the candidate material is very limited. Beside promoting the concept of critical-type nodal line, this work also provides an existing material to study the novel topological phase in future experiments.

\begin{acknowledgments}
This work is supported by the Special Foundation for Theoretical Physics Research Program of China (No. 11747152), the Natural Science Foundation of Hebei Province (No. E2016202383), Chongqing City Funds for Distinguished Young Scientists (No. cstc2014jcyjjq50003). One of the authors (G.D. Liu) acknowledges the financial support from Hebei Province Program for Top Young Talents.
\end{acknowledgments}

\begin{appendix}
\section{BAND STRUCTURE OF CaPd UNDER LATTICE DISTORTION AND ELEMENTAL DOPING}

\begin{figure}[t]
\includegraphics[width=8.5cm]{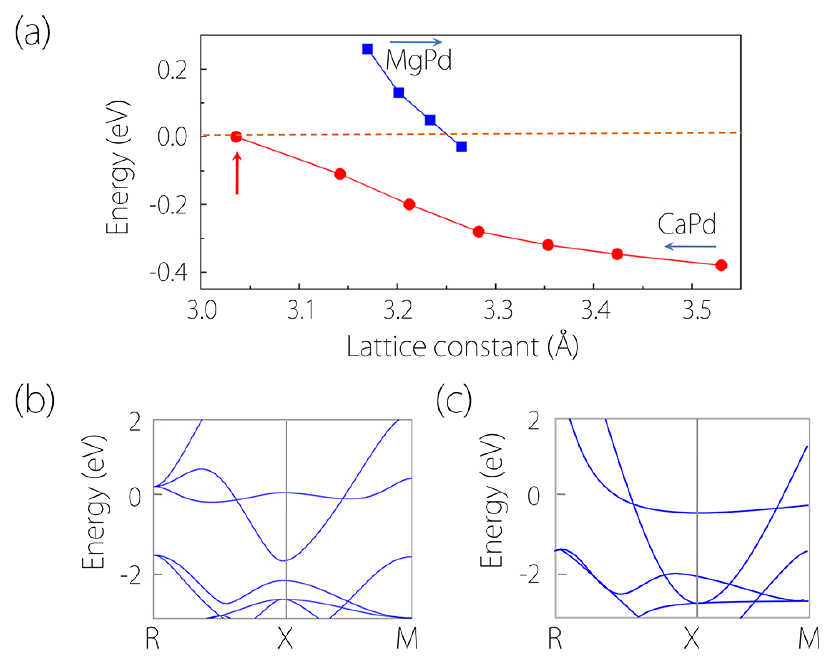}
\protect\caption{(a) The position of nodal line under different lattice constants for CaPd (red line) and MgPd (blue line). (b) Band structure of CaPd under 14\% lattice compression (at 3.03 \AA). (c) Band structure of CaPd under 75\% Mg doping (Ca$_{0.25}$Mg$_{0.75}$Pd).} \label{fig6}
\end{figure}

We find the position of nodal line in CaPd is quite sensitive with the lattice constants and elemental doping. Figure 6(a) shows the position of nodal line under different lattice constants of CaPd (here, we use the energy of crossing-A represents the position of the nodal line). One can find that the nodal line in CaPd moves towards the Fermi level upoun the lattice compression. The results show that, a 14\% lattice compression (at 3.03 \AA) finally makes the nodal line quite near the Fermi level, and corresponding band structure is shown in Fig. 6(b). Beside CaPd, we find MgPd also shows the critical-type band crossing near the Fermi level (at 0.22 eV above the Fermi level). Similarly, the nodal line in MgPd can be tuned to the Fermi level by a $\sim$3\% lattice expansion (at $\sim$3.24 \AA), as shown in Fig. 6(a). Starting from CaPd and MgPd, we find that a $\sim$75\% doping between them (Ca$_{0.25}$Mg$_{0.75}$Pd) can naturally give rise to a nodal line state near the Fermi level (-0.04 eV), as shown in Fig. 6(c). Importantly, the flat band in CaPd, which is crucial for the formation of critical-type nodal line, does not change much during lattice compression and Mg-doping [see Fig. 6(b) and (c)]. Moreover, their is no appearance of ferromagnetic order based on our DFT calculations during the lattice comprssion and elemental doping. Therefore, the critical-type nodal line state is promising to be realized in CaPd-based materials.

\end{appendix}

\end{document}